\documentstyle[12pt]{article}
 
\begin{document} 

\baselineskip 22pt 

\begin{center}
{\Large
\bf
Form Factors in $D$ Meson Decays}\\
\vspace{1.0cm}
Dae Sung Hwang${}^{(a)}$ and Do-Won Kim${}^{(b)}$\\
{\it{a: Department of Physics, Sejong University, Seoul 143--747,
Korea}}\\
{\it{b: Department of Physics, Kangnung National University,
Kangnung 210-702, Korea}}\\
\vspace{2.0cm}
{\bf Abstract}\\
\end{center}
\noindent
We study
the $d\Gamma / dq^2$ spectra and the branching fractions
of the $D$ meson exclusive semileptonic
decays with the lepton mass
effects into consideration.
We investigate their sensitivity to form factor models, and
find that the decays to a pseudoscalar meson and a lepton pair are
sensitive to the property of the form factor $F_1(q^2)$, and
those to a vector meson and a lepton pair
to the form factor $A_1(q^2)$.
We also analyze the experimental results of the branching fractions
${\cal{B}} (D^0\rightarrow K^- ({\rm or}\ \pi^-)\ \pi^+)$ and
${\cal{B}} (D^0\rightarrow K^- ({\rm or}\ \pi^-)\ e^+ \nu )$,
and show that it is implied that $F_1(q^2)$ is of dipole type, instead of
simple pole type which is commonly assumed.
\\

\vfill

\noindent
PACS codes: 13.20.Fc, 13.20.-v, 13.25.Ft, 14.40.Lb\\
Key words: D Meson, Semileptonic Decay, Lepton Mass Effect, Form Factor

\noindent
$^a$e-mail: dshwang@kunja.sejong.ac.kr\\
$^b$e-mail: dwkim@phys1.kangnung.ac.kr\\
\thispagestyle{empty}
\pagebreak

\baselineskip 22pt

\noindent
{\bf \large 1. Introduction}\\

The CP-violation phenomenon is discovered only in the
$K_L\rightarrow \pi\pi$ decay
and the charge asymmetry in the decay
$K_L\rightarrow {\pi}^{\pm} l^{\mp} \nu$
for more than 30 years.
The B-factories at KEK and SLAC are under construction for the discovery
of CP-violation in the $B$ meson system.
The mechanism of CP-violation through the complex phase of the
CKM three family mixing matrix \cite{ckm}
is presently considered standard for the CP-violation.
In order to measure the CKM matrix elements accurately, it is important to
know the hadronic form factors of the transition matrix elements reliably.
For the heavy to heavy transitions the heavy quark effective theory provides
good informations for the form factors.
However, for the heavy to light transitions
the understanding of the form factors is still limited and this fact hinders
the extractions of the CKM matrix elements from experimental results
significantly.
At the same time, we will be able to have important clues for the internal
structures of hadrons by knowing these form factors well.
Semileptonic decay processes are good sources for the
knowledge of the form factors both experimentally and theoretically,
and the lepton mass effects in heavy meson exclusive semileptonic decays
were studied by K\"orner and Schuler \cite{koerner}.

We derive the formulas for $d\Gamma /dq^2$ with non-zero lepton mass
in the forms which are efficient to study the form factor dependences.
This formula for the pseudoscalar to pseudoscalar transition was also given
by Khodjamirian et al. \cite{Khodjamirian}.
By using these formulas
we study the $d\Gamma /dq^2$ spectra and branching fractions
of the exclusive semileptonic $D$ meson decays:
$D^0\rightarrow K^{(*)-} e^+ \nu$,
$D^0\rightarrow K^{(*)-} \mu^+ \nu$,
$D^0\rightarrow \pi^- ({\rm or}\ \rho^-)\ e^+ \nu$
and $D^0\rightarrow \pi^- ({\rm or}\ \rho^-)\ \mu^+ \nu$.
In this anslysis we employ three models of form factors and show how the results
are influenced by the difference of form factors.
For the decays to a pseudoscalar meson and a lepton pair, the results are
sensitive to the property of the form factor $F_1(q^2)$.
For the decays to a vector meson and a lepton pair, the results are
sensitive to the form factor $A_1(q^2)$, and not to the other ones
($A_0(q^2)$, $A_2(q^2)$ and $V(q^2)$).
We also analyze the experimental results of the branching fractions
${\cal{B}} (D^0\rightarrow K^- ({\rm or}\ \pi^-)\ \pi^+)$ and
${\cal{B}} (D^0\rightarrow K^- ({\rm or}\ \pi^-)\ e^+ \nu )$,
and show that it is implied that the form factor
$F_1(q^2)$ of the $D$ to $K$ and the $D$ to $\pi$ transitions are of dipole type,
instead of simple pole type which is commonly assumed in the studies of the $D$
meson decays.
\\

\noindent
{\bf \large 2. Semileptonic Decays of Heavy Mesons}\\

{}From Lorentz invariance one finds the decomposition of the
hadronic matrix element
for pseudoscalar to pseudoscalar meson transition
in terms of hadronic form factors:
\begin{eqnarray}
& &<P(p)|J_\mu |P(P)>
\nonumber\\
&=&(P+p)_\mu f_+(q^2) + (P-p)_\mu f_-(q^2)
\nonumber\\
&=&
\Bigl( (P+p)_\mu 
-{M^2-m^2\over q^2}q_\mu \Bigr) \, F_1(q^2)
+{M^2-m^2\over q^2}\, q_\mu \, F_0(q^2),
\label{a1}
\end{eqnarray}
where $J_\mu = {\bar{q'}}\gamma_\mu (1-\gamma_5) q$.
We use the following notations:
$M$ represents initial meson mass,
$m$ final meson mass,
$m_l$ lepton mass,
$P$ initial meson momentum,
$p$ final meson momentum,
and $q_\mu =(P-p)_\mu$.
The form factors $F_1(q^2)$ and $F_0(q^2)$
correspond to $1^-$ and $0^+$ exchanges, respectively.
At $q^2=0$ we have the constraint
$F_1(0)=F_0(0)$,
since the hadronic matrix element in (\ref{a1}) is nonsingular
at this kinematic point.

The $q^2$ distribution of the semileptonic decay
$D^0\rightarrow K^- l^+ \nu$
is given in terms of the hadronic form factors
$F_1(q^2)$ and $F_0(q^2)$ as:
\begin{eqnarray}
&&{d\Gamma (D^0\rightarrow K^- l^+ \nu)\over dq^2}=
{G_F^2\over 24\pi^3}\, |V_{cs}|^2\, K(q^2)\,
(1-{m_l^2\over q^2})^2\times
\label{aa3}\\
&&
[\, (K(q^2))^2\, (1+{1\over 2}{m_l^2\over q^2})\, |F_1(q^2)|^2
\, +\, M^2\, (1-{m^2\over M^2})^2\, {3\over 8}\, {m_l^2\over q^2}
|F_0(q^2)|^2\, ]\, ,
\nonumber
\end{eqnarray}
where $K(q^2)$, momentum of the final meson in the $D$ meson rest frame,
is given by
\begin{equation}
K(q^2)={1\over 2M}\,
\Bigl( (M^2+m^2-q^2)^2
-4M^2m^2{\Bigr)}^{1\over 2} ,
\label{a3}
\end{equation}
and the physically allowed range of $q^2$ is given by
\begin{equation}
m_l^2\le q^2\le (M-m)^2.
\label{ab3}
\end{equation}
{}For $m_l=0$, (\ref{aa3}) is reduced to the
commonly used well-known formula:
\begin{equation}
{d\Gamma (D^0\rightarrow K^- l^+ \nu)\over dq^2}
={G_F^2\over 24\pi^3}\, |V_{cs}|^2\, (K(q^2))^3\,
|F_1(q^2)|^2,
\label{a2}
\end{equation}
and
$0\le q^2\le (M-m)^2$.
We note in (\ref{a2}) that only $F_1(q^2)$ contributes for $m_l=0$,
however, for $m_l\neq 0$ $F_0(q^2)$ also contributes as we can see
in (\ref{aa3}).

{}From Lorentz invariance one finds the decomposition of the
hadronic matrix element
for pseudoscalar to vector meson transition
in terms of hadronic form factors:
\begin{eqnarray}
& &<V(p)|J_\mu |P(P)>
\nonumber\\
&=&\varepsilon^{*\nu}(p)\Bigl( (M+m)g_{\mu\nu }A_1(q^2)
-2{P_\mu P_\nu \over M+m}A_2(q^2)
+{q_\mu P_\nu \over M+m}A_3(q^2)
\nonumber\\
&&+i\varepsilon_{\mu\nu\rho\sigma}{P^\rho p^\sigma \over M+m}V(q^2)\Bigr) ,
\label{b1}
\end{eqnarray}
where $\varepsilon_{0123}=1$ and
\begin{equation}
2mA_0(q^2)=(M+m)A_1(q^2)-{M^2-m^2+q^2\over M+m}A_2(q^2)
+{q^2\over M+m}A_3(q^2).
\label{b2}
\end{equation}
The form factors $V(q^2)$, $A_1(q^2)$, $A_2(q^2)$ and $A_0(q^2)$
correspond to $1^-$, $1^+$, $1^+$ and $0^-$ exchanges, respectively.
At $q^2=0$ we have the constraint
$2mA_0(0)=(M+m)A_1(0)-(M-m)A_2(0)$,
since the hadronic matrix element in (\ref{b1}) is nonsingular
at this kinematic point.

After a rather lengthy calculation,
the $q^2$ distribution of the semileptonic decay
$D^0\rightarrow K^{*-} l^+ \nu$
is given in terms of the hadronic form factors
$A_1(q^2)$, $A_2(q^2)$, $A_3(q^2)$ and $V(q^2)$ as \cite{hk}: 
\begin{eqnarray}
&&{d\Gamma (D^0\rightarrow K^{*-} l^+ \nu)\over dq^2}=
{G_F^2\over 32\pi^3}\, |V_{cs}|^2\, {1\over M^2} K(q^2)\,
(1-{m_l^2\over q^2})^2\times
\label{b4}\\
&&\{ |A_1(q^2)|^2{(M+m)^2\over m^2}[{1\over 3}(MK)^2(1-{m_l^2\over q^2})
+q^2m^2+(MK)^2{m_l^2\over q^2}+{1\over 2}m^2m_l^2]
\nonumber\\
&&\ +{\rm Re}(A_1(q^2)A_2^*(q^2))[
-{M^2-m^2-q^2\over m^2}
[{2\over 3}(MK)^2(1-{m_l^2\over q^2})
+2(MK)^2{m_l^2\over q^2}
\nonumber\\
&&\qquad\qquad\qquad\qquad +{1\over 2}(M^2+m^2-q^2)m_l^2]
+(M^2-m^2+q^2)m_l^2]
\nonumber\\
&&\ +|A_2(q^2)|^2{1\over (M+m)^2m^2}(MK)^2[{4\over 3}(MK)^2(1-{m_l^2\over q^2})
+4(MK)^2{m_l^2\over q^2}+2M^2m_l^2]
\nonumber\\
&&\ +|V(q^2)|^2{q^2\over (M+m)^2}[{8\over 3}(MK)^2(1-{m_l^2\over q^2})
+4(MK)^2{m_l^2\over q^2}]
\nonumber\\
&&\ +|A_3(q^2)|^2{q^2\over (M+m)^2m^2}{1\over 2}(MK)^2m_l^2
\nonumber\\
&&\ -{\rm Re}(A_3(q^2)A_2^*(q^2)){1\over (M+m)^2m^2}(M^2-m^2+q^2)(MK)^2m_l^2
\nonumber\\
&&\ +{\rm Re}(A_3(q^2)A_1^*(q^2)){1\over m^2}(MK)^2m_l^2\} .
\nonumber
\end{eqnarray}
When we take $m_l\to 0$ in (\ref{b4}), it agrees with the formula for $m_l=0$
given in Refs. \cite{GS, ukqcdD51}:
\begin{equation}
{d\Gamma (D^0\rightarrow K^{*-} l^+ \nu)\over dq^2}=
{G_F^2\over 96\pi^3}\, |V_{cs}|^2\,{q^2\over M^2} K(q^2)\, 
(|H^+(q^2)|^2+|H^-(q^2)|^2+|H^0(q^2)|^2),
\end{equation}
where
\begin{eqnarray}
H^0(q^2)&=&{-1\over 2m{\sqrt{q^2}}}\Bigl(
(M^2-m^2-q^2)(M+m)A_1(q^2)-{4M^2K^2\over M+m}A_2(q^2)\Bigr)\, ,
\label{GS}\\
H^{\pm}(q^2)&=&-\Bigl( (M+m)A_1(q^2)\mp {2MK\over M+m}V(q^2)\Bigr)\, .
\nonumber
\end{eqnarray}
In the case of the $B$ to $D$ meson (heavy to heavy) transition,
the heavy quark effective theory (HQET) gives the useful relations between the
relevant form factors \cite{iw}:
\begin{eqnarray}
F_1(q^2)&=&V(q^2)\ =\ A_0(q^2)\ =\ A_2(q^2)\ =\
{M+m\over 2{\sqrt{Mm}}}\,\, {\cal{F}} (y),
\label{c1}\\
F_0(q^2)&=&A_1(q^2)\ =\
{2{\sqrt{Mm}}\over M+m}\,\, {y+1\over 2}\,\, {\cal{F}} (y),
\nonumber
\end{eqnarray}
where $y=(M^2+m^2-q^2)/(2Mm)=E_{D^{(*)}}/m$ ($E_{D^{(*)}}$ is
the energy of $D^{(*)}$ meson in the $B$ meson rest frame),
and ${\cal{F}} (y)$ is a form factor which becomes the Isgur-Wise function
in the infinite heavy quark mass limit.
When we use the relations (\ref{c1}),
for $m_l=0$ the formula (\ref{b4}) becomes
the well-known formula for the $B$ to $D^*$ transition:
\begin{eqnarray}
{d\Gamma ({\bar{B}}^0\rightarrow D^{*+} l^- {\bar{\nu}})\over dq^2}&=&
{G_F^2\over 48\pi^3}\, |V_{cb}|^2\, m^3\, (M-m)^2\, {\sqrt{y^2-1}}\,
(y+1)^2\times
\nonumber\\
&&\{ 1+{4y\over y+1}\, {1-2yr+r^2\over (1-r)^2} \} \,\, 
({\cal{F}}_{D^*} (y))^2,
\label{b7}
\end{eqnarray}
where $r=m/M$.
\\

\noindent
{\bf \large 3. $D^0\rightarrow K^{(*)-} l^+ \nu$}\\

{}For the form factors concerned with the exclusive semileptonic decays of $D$
meson, we can not use the relations (\ref{c1}) of the HQET.
Therefore, in the study of $D$ meson decays
we use models for form factors.
The pole-dominance idea suggests the following $q^2$ dependence of the
form factors \cite{wsb}:
\begin{equation}
f_i(q^2)=f_i(0)\, {1\over (1-{q^2\over m_{f_i}^2})^{n_{f_i}}},
\label{a4}
\end{equation}
where $n_{f_i}$ and $m_{f_i}$ are corresponding
power and pole mass of the form factors $f_i(q^2)$, respectively.
The WSB model \cite{wsb} adopts $n_{f_i}=1$.
However, the exact values of $n_{f_i}$ are not known.
The relations (\ref{c1}) of the HQET gives the following
approximate relation among the powers of the form factors
for the heavy to heavy transitions:
\begin{equation}
n_{F_1}=n_{V}=n_{A_0}=n_{A_2}=n_{F_0}+1=n_{A_1}+1.
\label{hqetrn}
\end{equation}
Non-perturbative analysis of QCD \cite{hl} suggests the same relation as
(\ref{hqetrn}) for the form factors of the heavy to light transitions.
The lattice calculations also show
that the form factors $F_1$, $V$ and $A_0$ are more rapidly increasing
functions of $q^2$ than the form factors $F_0$ and $A_1$ \cite{ukqcdD51, ukqcd},
which favors the relation (\ref{hqetrn}).
Therefore, we will adopt two other models incorporating the relation
(\ref{hqetrn}), as well as the WBS model which assumes $n_{f_i}=1$ \cite{wsb},
for the study of the exclusive semileptonic decays of $D$ meson.
We organize in Table 1 the values of the powers $n_{f_i}$ of the three models
which we use in this work.
{}For the values of the pole masses and those of the form factors at $q^2=0$,
we use the values organized in Table 2 and 3, which were
given by Wirbel, Stech and Bauer \cite{wsb}.
Their precise values are not known and they should be different for each
of the three models.
However, their exact values are not crucial for our work of clarifying
the model dependences
of the $d\Gamma / dq^2$ spectra and the branching fractions.

{}For $D^0 \rightarrow K^- l^+ \nu$,
we use the formula (\ref{aa3}) with non-zero lepton mass,
instead of the commonly used formula (\ref{a2}) which is true for zero lepton
mass.
The obtained $d\Gamma (D^0 \rightarrow K^- l^+ \nu ) / dq^2$ spectrum
and branching fractions
are presented in Figure 1 and Table 4, for each of the three models we
adopt in this work:
WSB, Model I and Model II explained in Table 1.
We find that the shape of spectrum of Model II is different from
those of WSB and Model I in Figure 1.
That is, the value of $n_{F_1}$ determines the shape of spectrum.
The experimental result of the E687 Collaboration for this spectrum was given in
Fig. 3 (a) of Ref. \cite{E68795}, and the shape of their spectrum favors
Model II better than WSB and Model I.
In the experimental extraction of the value of $F_1^{DK}(0)$,
the simple pole of the form factors has been commonly assumed \cite{pdg94, ryd}.
Under this assumption, $F_1^{DK}(0)=0.75\pm 0.02\pm 0.02$ was extracted
\cite{pdg94} from
the experimentally measured branching fraction
${\cal{B}}(D^0 \rightarrow K^- e^+ \nu)=(3.68\pm 0.21)\times 10^{-2}$.
(In Ref. \cite{ryd}, $F_1^{DK}(0)=0.76\pm 0.03$ was presented.)
However, if we assume in this extraction
Model II which has the dipole form factor for $F_1(q^2)$,
we would get
\begin{equation}
F_1^{DK}(0)=0.75\times {\sqrt{3.49\times 10^{-2}\over 4.78\times 10^{-2}}}
=0.75\times 0.85=0.64
\label{f1value}
\end{equation}
for the mean value.
In (\ref{f1value}) we used our results in Table 4 of the branching fraction
${\cal{B}}(D^0 \rightarrow K^- e^+ \nu)$ for WSB ($n_{F_1}=1$)
and Model II ($n_{F_1}=2$),
which were obtained by using the same value of $F_1^{DK}(0)$.
That is, the experimentally extracted value of $F_1^{DK}(0)$ is much
dependent on the form factor models used in the analysis.
In reality, it is not yet established which type of form factors is the right one.

{}For $D^0 \rightarrow K^{*-} l^+ \nu$,
we use the formula (\ref{b4}) with non-zero lepton mass.
The obtained spectra and branching fractions are presented in Figure 2
and Table 5.
We find that the results of WSB and Model II are almost the same, and
they are significantly different from the results of Model I.
This fact implies that the value of $n_{A_1}$ mainly determines the spectra
and branching fractions of $D^0 \rightarrow K^{*-} l^+ \nu$.
\\

\noindent
{\bf \large 4. $D^0\rightarrow \pi^- ({\rm or}\ \rho^- )\ l^+ \nu$}\\

{}For $D^0 \rightarrow \pi^- l^+ \nu$,
we use the formula (\ref{aa3}) with the replacement of $V_{cs}$ by $V_{cd}$.
The obtained spectra and
branching fractions are presented in Fig. 3 and Table 6.
We find that the branching fractions of Model II in Fig. 3
are about twice those of WSB and Model I.
Therefore, in case that we determine the
value of $F_1^{D\pi}(0)$ from an experimentally measured branching fraction
of $D^0 \rightarrow \pi^- l^+ \nu$,
the value of $F_1^{D\pi}(0)$ determined with Model II will be about
$1/{\sqrt{2}}$ times its value
determined with WSB or Model I.
From Table 4 and 6, we also find that the ratio
${\cal{B}}(D^0 \rightarrow \pi^- l^+ \nu)/
{\cal{B}}(D^0 \rightarrow K^{-} l^+ \nu)$
from Model II is pretty bigger than that from WSB or Model I, but its present
experimental result $0.101\pm 0.020\pm 0.003$ \cite{E68796} can not discriminate
them yet.

{}For $D^0 \rightarrow \rho^- l^+ \nu$,
we use the formula (\ref{b4}) with the replacement of $V_{cs}$ by $V_{cd}$.
The obtained spectra and
branching fractions are presented in Fig. 4 and Table 7.
We find in Fig. 4 and Table 7 that the results of WSB and Model II are almost
the same,
and they are much different from the results of Model I.
Therefore, like the $D^0 \rightarrow K^{*-} l^+ \nu$ case, the property of
the form factor $A_1(q^2)$ mainly
determines the spectra and branching fractions,
and the results are not sensitive to the other form factors
($A_0(q^2)$, $A_2(q^2)$ and $V(q^2)$).
\\

\noindent
{\bf \large 5. Implications of Experimental Results}\\

In this section we compare the exclusive semileptonic decays and the two-body
hadronic decays.
We start by recalling the relevant effective weak Hamiltonian for the two-body
hadronic decay $D^0\rightarrow K^- \pi^+$:
\begin{equation}
{\cal H}_{\rm eff}={G_F\over {\sqrt{2}}}V_{cs}^*V_{ud}
[C_1(\mu ){\cal O}_1+C_2(\mu ){\cal O}_2]\ +\ {\rm H.C.},
\label{b1}
\end{equation}
\begin{equation}
{\cal O}_1=({\bar{u}}\Gamma^\rho d)({\bar{s}}\Gamma_\rho c),\quad
{\cal O}_2=({\bar{s}}\Gamma^\rho d)({\bar{u}}\Gamma_\rho c),
\label{b2}
\end{equation}
where $G_F$ is the Fermi coupling constant, $V_{cs}$ and $V_{ud}$
are corresponding Cabibbo-Kobayashi-Maskawa (CKM) matrix elements
and $\Gamma_\rho = \gamma_\rho (1-\gamma_5)$.
The Wilson coefficients $C_1(\mu )$ and $C_2(\mu )$ incorporate
the short-distance effects arising from the renormalization of
${\cal H}_{\rm eff}$ from $\mu =m_W$ to $\mu =O(m_c)$.
By using the Fierz transformation under which $V-A$ currents remain
$V-A$ currents, we get the following equivalent forms:
\begin{eqnarray}
C_1{\cal O}_1+C_2{\cal O}_2&=&
(C_1+{1\over N_c}C_2){\cal O}_1
+C_2({\bar{s}}\Gamma^\rho T^ad)({\bar{u}}\Gamma_\rho T^ac)
\nonumber\\
&=&(C_2+{1\over N_c}C_1){\cal O}_2
+C_1({\bar{u}}\Gamma^\rho T^ad)({\bar{s}}\Gamma_\rho T^ac),
\label{b3}
\end{eqnarray}
where $N_c=3$ is the number of colors and $T^a$'s are $SU(3)$ color
generators.
The second terms in (\ref{b3}) involve color-octet currents.
In the factorization assumption, these terms are neglected and
${\cal H}_{\rm eff}$ is rewritten in terms of ``factorized hadron
operators'' \cite{wsb}:
\begin{equation}
{\cal H}_{\rm eff}={G_F\over {\sqrt{2}}}V_{cs}^*V_{ud}
\Bigl( a_1[{\bar{u}}\Gamma^\rho d]_H[{\bar{s}}\Gamma_\rho c]_H
+a_2[{\bar{s}}\Gamma^\rho d]_H[{\bar{u}}\Gamma_\rho c]_H\Bigr)
\ +\ {\rm H.C.},
\label{bb4}
\end{equation}
where the subscript $H$ stands for $hadronic$ implying that the
Dirac bilinears inside the brackets be treated as interpolating
fields for the mesons and no further Fierz-reordering need be done.
The phenomenological parameters $a_1$ and $a_2$ are related to
$C_1$ and $C_2$ by
$a_1=C_1+{1\over N_c}C_2$ and
$a_2=C_2+{1\over N_c}C_1$.
The numerical values of $a_1$ and
$a_2$ for $D$ meson decays are given by \cite{NS}
\begin{equation}
a_1=1.10\pm 0.05\, ,\quad a_2=-0.49\pm 0.04\, .
\label{b6}
\end{equation}

For the two body decay, in the rest frame of initial meson
the differential decay rate is given by
\begin{equation}
d\Gamma ={1\over 32\pi^2}|{\cal M}|^2
{|{\bf p}_1|\over M^2}d\Omega ,
\label{b7}
\end{equation}
\begin{equation}
|{\bf p}_1|=
{[(M^2-(m_1+m_2)^2)(M^2-(m_1-m_2)^2)]^{{1\over 2}}\over 2M},
\label{b8}
\end{equation}
where $M$ is the initial meson mass, $m_1$ and $m_2$ the final meson masses, and
${\bf p}_1$ the momentum of one of the
final mesons in the initial meson rest frame.
By using (\ref{a1}), (\ref{bb4}) and
$<0|\Gamma_\mu |\pi^-(q)>=iq_\mu f_{\pi^-}$,
(\ref{b7}) gives the following formula for the branching ratio of the
process $D^0\rightarrow K^- \pi^+$:
\begin{eqnarray}
&&{\cal{B}} (D^0\rightarrow K^- \pi^+)
=({G_Fm_D^2\over {\sqrt{2}}})^2\,
|V_{ud}|^2\, {1\over 8\pi }\, {m_D\over {\Gamma}_D}\, a_1^2\,
{f_{\pi}^2 \over m_D^2}\, |V_{cs}\, F_0^{DK}(m_{\pi}^2)|^2
\nonumber\\
& &\qquad\times \Bigl( 1-{m_K^2\over m_D^2}{\Bigr)}^2\,
{1\over 2}\,
[\Bigl( 1-({m_K+m_{\pi}\over m_D})^2{\Bigr)}
\Bigl( 1-({m_K-m_{\pi}\over m_D})^2{\Bigr)}]^{1\over 2}\, .
\label{b11}
\end{eqnarray}

On the other hand, from (\ref{a2}) and (\ref{a4})
the branching ratio ${\cal{B}} (B^0\rightarrow K^- e^+ \nu )$
is given by
\begin{eqnarray}
{\cal{B}} (D^0\rightarrow K^- e^+ \nu )&=&
({G_Fm_D^2\over {\sqrt{2}}})^2\, {m_D\over {\Gamma}_D}\,
{2\over 192 {\pi}^3}\,
|V_{cs}\, F_1^{DK}(0)|^2\times I^{DK}\, ,
\label{a6}
\end{eqnarray}
where the dimensionless integral $I^{DK}$ is given by
\begin{equation}
I^{DK}=
\int_0^{(1-{m_{K}\over m_D})^2} dx
{\Bigl( (1+{m_{K}^2\over m_D^2}-x)^2
-4{m_{K}^2\over m_D^2}{\Bigr)}^{3\over 2}\over
\Bigl( 1-{m_D^2\over m_{F_1}^2}x{\Bigr)}^{2n_{F_1}}}.
\label{a7}
\end{equation}
In the above, we neglected the electron mass.
From (\ref{b11}) and (\ref{a7}) we have
\begin{eqnarray}
&&{{\cal{B}} (D^0\rightarrow K^- \pi^+) \over
{\cal{B}} (D^0\rightarrow K^- e^+ \nu )}\, =\, 
6\pi^2\, |V_{ud}|^2\, {f_{\pi}^2 \over m_D^2}\, 
\Bigl( 1-{m_K^2\over m_D^2}{\Bigr)}^2
\nonumber\\
&&\qquad\times 
[\Bigl( 1-({m_K+m_{\pi}\over m_D})^2{\Bigr)}
\Bigl( 1-({m_K-m_{\pi}\over m_D})^2{\Bigr)}]^{1\over 2}\,
{|V_{cs}\, F_0^{DK}(m_{\pi}^2)|^2\over
|V_{cs}\, F_0^{DK}(0)|^2}\ {a_1^2\over I^{DK}}
\nonumber\\
&&\qquad\qquad
=0.225\times {a_1^2\over I^{DK}}\, ,
\label{zz1}
\end{eqnarray}
where we used the fact $F_0^{DK}(m_{\pi}^2)\simeq F_0^{DK}(0)$ and
the following experimetal values \cite{rpp}:
$m_D=m_{D^0}=1.8646\pm 0.0005$ GeV, $m_K=m_{K^-}=493.677\pm 0.013$ MeV,
$m_{\pi}=m_{\pi^+}=139.56995\pm 0.00035$ MeV,
$f_{\pi}=f_{\pi^+}=131.74\pm 0.15$ MeV
and $V_{ud}=0.9753\pm 0.0008$.

When we use the experimental results
${\cal{B}} (D^0\rightarrow K^- \pi^+)=3.85\pm 0.09\ \%$ and
${\cal{B}} (D^0\rightarrow K^- e^+ \nu )=3.66\pm 0.18\ \%$, (\ref{zz1}) gives
\begin{eqnarray}
I^{DK}[{\rm Expt.}]&=&0.213\ (1\pm 0.054)\ a_1^2
\ =\ 0.258\ (1\pm 0.054)\ (1\pm 0.091)
\nonumber\\
&=&0.258\ (1\pm 0.106)\ =\ 0.231\sim 0.286\, ,
\label{zz2}
\end{eqnarray}
where we used the value of $a_1$ given in (\ref{b6}).
On the other hand,
when we calculate $I^{DK}$ directly from (\ref{a7}) with $m_{F_1}=2.11$ GeV
geven in Table 2, we obtain the following results:
\begin{eqnarray}
I^{DK}[n_{F_1}=1]&=&0.195\qquad {\rm for}\ \ n_{F_1}=1,
\nonumber\\
I^{DK}[n_{F_1}=2]&=&0.267\qquad {\rm for}\ \ n_{F_1}=2.
\label{zz3}
\end{eqnarray}
From (\ref{zz2}) and (\ref{zz3}), we find that the experimental results
of ${\cal{B}} (D^0\rightarrow K^- \pi^+)$ and
${\cal{B}} (D^0\rightarrow K^- e^+ \nu )$
imply $n_{F_1}=2$.

In the same way as the above, for the $D$ to $\pi$ transition we get the
formula
\begin{eqnarray}
&&{{\cal{B}} (D^0\rightarrow \pi^- \pi^+) \over
{\cal{B}} (D^0\rightarrow \pi^- e^+ \nu )}\, =\,
6\pi^2\, |V_{ud}|^2\, {f_{\pi}^2 \over m_D^2}\,
\Bigl( 1-{m_{\pi}^2\over m_D^2}{\Bigr)}^2
\nonumber\\
&&\qquad\times
[\Bigl( 1-({m_{\pi}+m_{\pi}\over m_D})^2{\Bigr)}]^{1\over 2}\,
{|V_{ud}\, F_0^{DK}(m_{\pi}^2)|^2\over
|V_{ud}\, F_0^{DK}(0)|^2}\ {a_1^2\over I^{D\pi}}
\nonumber\\
&&\qquad\qquad
=0.275\times {a_1^2\over I^{D\pi}}\, ,
\label{zz4}
\end{eqnarray}
where we used the fact $F_0^{D\pi}(m_{\pi}^2)\simeq F_0^{D\pi}(0)$, and
the dimensionless integral $I^{D\pi}$ is given by
\begin{equation}
I^{D\pi}=
\int_0^{(1-{m_{\pi}\over m_D})^2} dx
{\Bigl( (1+{m_{\pi}^2\over m_D^2}-x)^2
-4{m_{\pi}^2\over m_D^2}{\Bigr)}^{3\over 2}\over
\Bigl( 1-{m_D^2\over m_{F_1}^2}x{\Bigr)}^{2n_{F_1}}}.
\label{a7b}
\end{equation}

When we use the experimental results
${\cal{B}} (D^0\rightarrow \pi^- \pi^+)=(1.53\pm 0.09)\times 10^{-3}$ and
${\cal{B}} (D^0\rightarrow \pi^- e^+ \nu )=(3.7\pm 0.6)\times 10^{-3}$,
(\ref{zz4}) gives
\begin{eqnarray}
I^{D\pi}[{\rm Expt.}]&=&0.665\ (1\pm 0.173)\ a_1^2
\ =\ 0.804\ (1\pm 0.173)\ (1\pm 0.091)
\nonumber\\
&=&0.804\ (1\pm 0.195)\ =\ 0.648\sim 0.961\, ,
\label{zz5}
\end{eqnarray}
where we used again the value of $a_1$ given in (\ref{b6}).
When we calculate $I^{D\pi}$ directly from (\ref{a7}) with $m_{F_1}=2.01$ GeV
geven in Table 2, we obtain the following results:
\begin{eqnarray}
I^{D\pi}[n_{F_1}=1]&=&0.385\qquad {\rm for}\ \ n_{F_1}=1,
\nonumber\\
I^{D\pi}[n_{F_1}=2]&=&0.783\qquad {\rm for}\ \ n_{F_1}=2.
\label{zz6}
\end{eqnarray}
The results in (\ref{zz6}) show that the value of $I^{D\pi}$ is more sensitive
to $n_{F_1}$ than that of $I^{DK}$.
From (\ref{zz5}) and (\ref{zz6}), we find that the experimental results
of ${\cal{B}} (D^0\rightarrow \pi^- \pi^+)$ and
${\cal{B}} (D^0\rightarrow \pi^- e^+ \nu )$
imply $n_{F_1}=2$.
\\

\noindent
{\bf \large 6. Conclusion}\\

We studied the $D$ meson exclusive semileptonic
decays with the lepton mass effects into consideration,
and investigated their sensitivity to form factor models.
The results show that the decays to a pseudoscalar meson and a lepton pair are
sensitive to the property of the form factor $F_1(q^2)$, and
those to a vector meson and a lepton pair are
sensitive to the form factor $A_1(q^2)$ and not to the other ones
($A_0(q^2)$, $A_2(q^2)$ and $V(q^2)$).
Concerned with the experimental extraction of the form factor values at $q^2=0$,
$F_1^{DK}(0)=0.75\pm 0.02\pm 0.02$ has been obtained from the experimental result
${\cal{B}}(D^0 \rightarrow K^- e^+ \nu)=(3.68\pm 0.21)\times 10^{-2}$ by
assuming $n_{F_1}=1$ \cite{pdg94}, however, its mean value is modified to $0.64$ if
$n_{F_1}=2$ is adopted.
In reality, the value of $n_{F_1}$ is not known presently.
The experimental extraction of the reliable value of $F_1^{DK}(0)$ is very
important for the determinations of the fundamental parameters, and also for
the reason that the experimentally extracted value of $F_1^{DK}(0)$ is compared
with its values obtained by different theoretical calculations such as
quark models, lattice QCD, and QCD sum rules \cite{pdg94, ryd}.
We emphasized
that this extraction is very much dependent on the value of $n_{F_1}$.
In this context, we also
analyzed the experimental results of the branching fractions
${\cal{B}} (D^0\rightarrow K^- ({\rm or}\ \pi^-)\ \pi^+)$ and
${\cal{B}} (D^0\rightarrow K^- ({\rm or}\ \pi^-)\ e^+ \nu )$,
and showed that it is implied that the form factor
$F_1(q^2)$ of the $D$ to $K$ and that of
the $D$ to $\pi$ transitions are of dipole type ($n_{F_1}=2$),
instead of simple pole type ($n_{F_1}=1$)
which is commonly assumed in the studies of the $D$ meson decays.
We think that careful studies on the form factors are
important not only for the extractions of the CKM matrix elements
but also for the understanding of the internal structures of hadrons.
\\

\vspace*{1.0cm}

\noindent
{\em Acknowledgements} \\
\noindent
This work was supported
by Non-Directed-Research-Fund,
Korea Research Foundation 1997,
by the Basic Science Research Institute Program,
Ministry of Education, Project No. BSRI-97-2414,
by the Korea Science and Engineering Foundation,
Grant 985-0200-002-2,
and by the Research Fund of
Kangnung National University 1997.
\\

\pagebreak

\pagebreak

\begin{table}[!h]
\begin{center}
\begin{tabular}{|c|c|c|c|c|c|c|}   \hline
&$n_{F_1}$&$n_{F_0}$&$n_V$&$n_{A_1}$&$n_{A_2}$&$n_{A_0}$
\\   \hline
WSB(pole/pole)       &1&1&1&1&1&1\\
Model I(pole/const.) &1&0&1&0&1&1\\
Model II(dipole/pole)&2&1&2&1&2&2\\
\hline
\end{tabular}
\end{center}
\caption{The values of the power of pole for the three models
used in this paper.}
\end{table}
\begin{table}[!h]
\begin{center}
\begin{tabular}{|c|c|c|c|c|}   \hline
Current &$m(0^-)$&$m(1^-)$&$m(0^+)$&$m(1^+)$\\
Relevant Form Factors&$A_0$&$F_1\ V$&$F_0$&$A_1\ A_2$
\\   \hline
${\bar{s}}c$&1.97&2.11&2.60&2.53\\
${\bar{d}}c$&1.87&2.01&2.47&2.42\\
\hline
\end{tabular}
\end{center}
\caption{The values of pole masses (GeV) used in numerical calculations.}
\end{table}
\begin{table}[!h]
\begin{center}
\begin{tabular}{|c|c|c|c|c|c|}   \hline
&${F_1(0)}={F_0(0)}$&$V(0)$&${A_1(0)}$&${A_2(0)}$&${A_0(0)}$
\\   \hline
$D\to K^{(*)}$   &0.762&1.226&0.880&1.147&0.733\\
$D\to\pi (\rho )$&0.692&1.225&0.775&0.923&0.669\\
\hline
\end{tabular}
\end{center}
\caption{The values of the form factors at $q^2=0$
used in numerical calculations.}
\end{table}

\pagebreak

\begin{table}[!h]
\begin{center}
\begin{tabular}{|c|c|c|c|}   \hline
&${\cal{B}}(D^0\to K^- e^+ \nu)$
&${\cal{B}}(D^0\to K^- \mu^+ \nu)$
&${\cal{B}}( e^+ ) :
{\cal{B}}( \mu^+ )$
\\   \hline
WSB &$3.49\times 10^{-2}$&$3.41\times 10^{-2}$&
1\, :\, 0.98\\
Model I &$3.49\times 10^{-2}$&$3.38\times 10^{-2}$&
1\, :\, 0.97\\
Model II&$4.78\times 10^{-2}$&$4.67\times 10^{-2}$&
1\, :\, 0.98\\
\hline
\end{tabular}
\end{center}
\caption{The obtained branching fractions and their ratios for
$D^0\to K^- l^+ \nu$.}
\end{table}
\begin{table}[!h]
\begin{center}
\begin{tabular}{|c|c|c|c|}   \hline
&${\cal{B}}(D^0\to K^{*-} e^+ \nu)$
&${\cal{B}}(D^0\to K^{*-} \mu^+ \nu)$
&${\cal{B}}( e^+ ) :
{\cal{B}}( \mu^+ )$
\\   \hline
WSB &$3.88\times 10^{-2}$&$3.68\times 10^{-2}$&
1\, :\, 0.95\\
Model I &$3.28\times 10^{-2}$&$3.10\times 10^{-2}$&
1\, :\, 0.95\\
Model II&$3.85\times 10^{-2}$&$3.66\times 10^{-2}$&
1\, :\, 0.95\\
\hline
\end{tabular}
\end{center}
\caption{The obtained branching fractions and their ratios for
$D^0\to K^{*-} l^+ \nu$.}
\end{table}
\begin{table}[!h]
\begin{center}
\begin{tabular}{|c|c|c|c|}   \hline
&${\cal{B}}(D^0\to \pi e^+ \nu)$
&${\cal{B}}(D^0\to \pi \mu^+ \nu)$
&${\cal{B}}( e^+ ) :
{\cal{B}}( \mu^+ )$
\\   \hline
WSB &$2.92\times 10^{-3}$&$2.88\times 10^{-3}$&
1\, :\, 0.99\\
Model I &$2.92\times 10^{-3}$&$2.86\times 10^{-3}$&
1\, :\, 0.98\\
Model II&$5.94\times 10^{-3}$&$5.86\times 10^{-3}$&
1\, :\, 0.99\\
\hline
\end{tabular}
\end{center}
\caption{The obtained branching fractions and their ratios for
$D^0\to \pi^- l^+ \nu$.}
\end{table}
\begin{table}[!h]
\begin{center}
\begin{tabular}{|c|c|c|c|}   \hline
&${\cal{B}}(D^0\to \rho e^+ \nu)$
&${\cal{B}}(D^0\to \rho \mu^+ \nu)$
&${\cal{B}}( e^+ ) :
{\cal{B}}( \mu^+ )$
\\   \hline
WSB &$2.77\times 10^{-3}$&$2.66\times 10^{-3}$&
1\, :\, 0.96\\
Model I &$2.19\times 10^{-3}$&$2.10\times 10^{-3}$&
1\, :\, 0.96\\
Model II&$2.77\times 10^{-3}$&$2.66\times 10^{-3}$&
1\, :\, 0.96\\
\hline
\end{tabular}
\end{center}
\caption{The obtained branching fractions and their ratios for
$D^0\to \rho^- l^+ \nu$.}
\end{table}

\pagebreak

\noindent
{\large\bf
Figure Captions}\\

\noindent
Fig. 1. $(1/\Gamma_{tot})(d\Gamma/dq^2)$ of
(a) $D^0 \rightarrow K^{-} e^+ \nu$ and
(b) $D^0 \rightarrow K^{-} \mu^+ \nu$ for three models:
solid line for WSB, dashed for Model I,
and dotted for Model II.
Solid and dashed lines almost overlap.
\\

\noindent
Fig. 2. $(1/\Gamma_{tot})(d\Gamma/dq^2)$ of
$D^0 \rightarrow K^{*-} e^+ \nu$ and
$D^0 \rightarrow K^{*-} \mu^+ \nu$ for three models:
solid line for WSB, dashed for Model I,
and dotted for Model II.
Solid and dotted lines almost overlap.
\\

\noindent
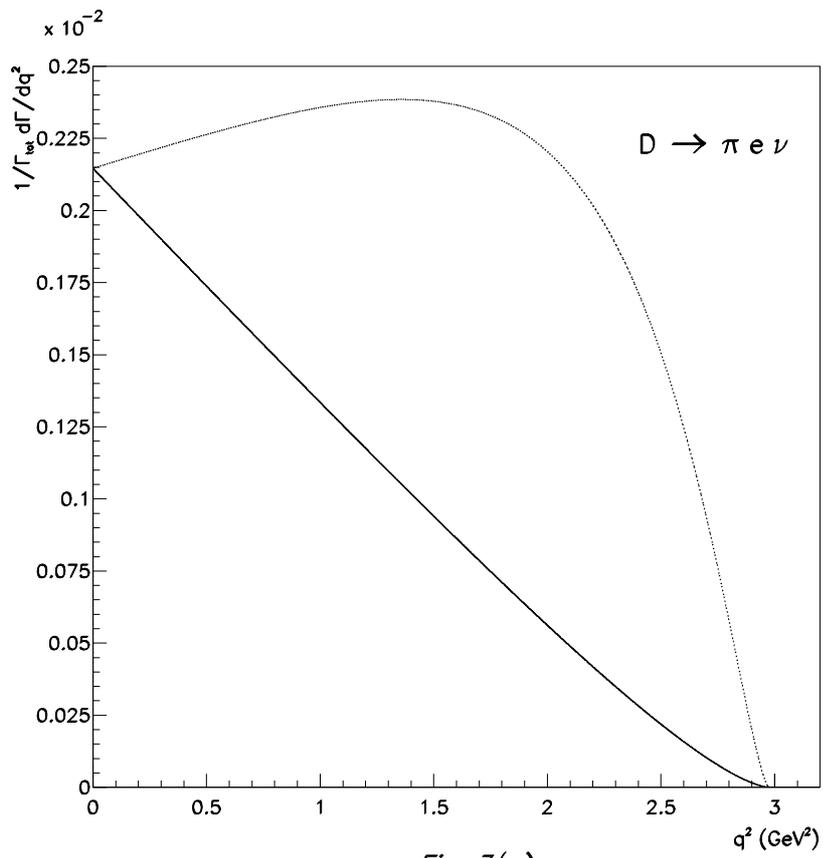
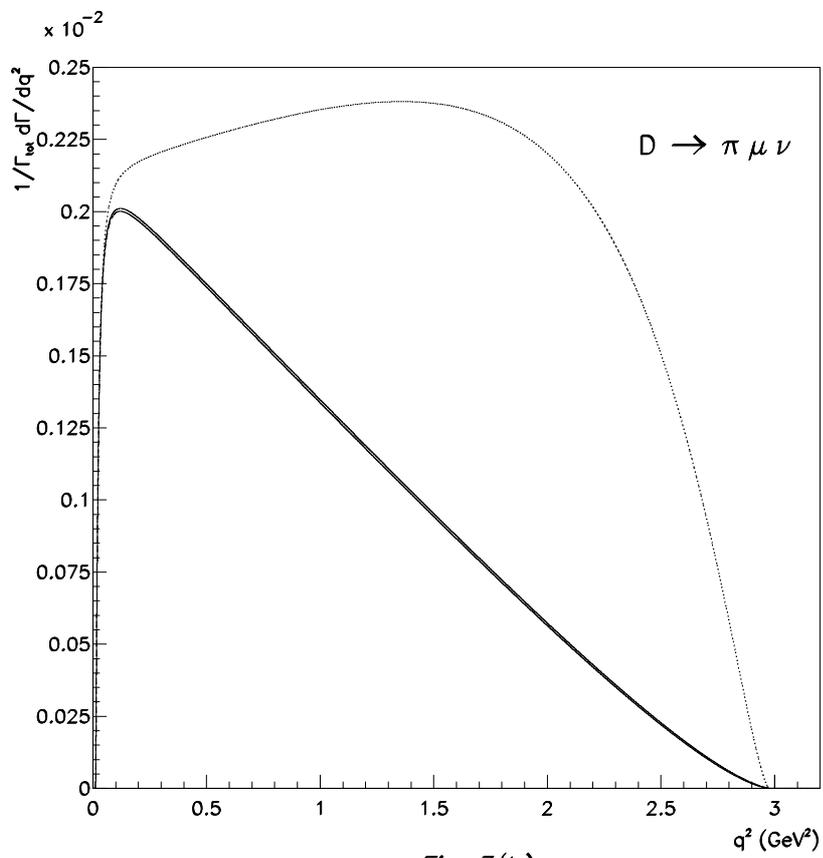
Fig. 3. $(1/\Gamma_{tot})(d\Gamma/dq^2)$ of
(a) $D^0 \rightarrow \pi^{-} e^+ \nu$ and
(b) $D^0 \rightarrow \pi^{-} \mu^+ \nu$ for three models:
solid line for WSB, dashed for Model I,
and dotted for Model II.
Solid and dashed lines almost overlap.
\\

\noindent
Fig. 4. $(1/\Gamma_{tot})(d\Gamma/dq^2)$ of
$D^0 \rightarrow \rho^{-} e^+ \nu$ and 
$D^0 \rightarrow \rho^{-} \mu^+ \nu$ for three models:
solid line for WSB, dashed for Model I,
and dotted for Model II.
Solid and dotted lines almost overlap.

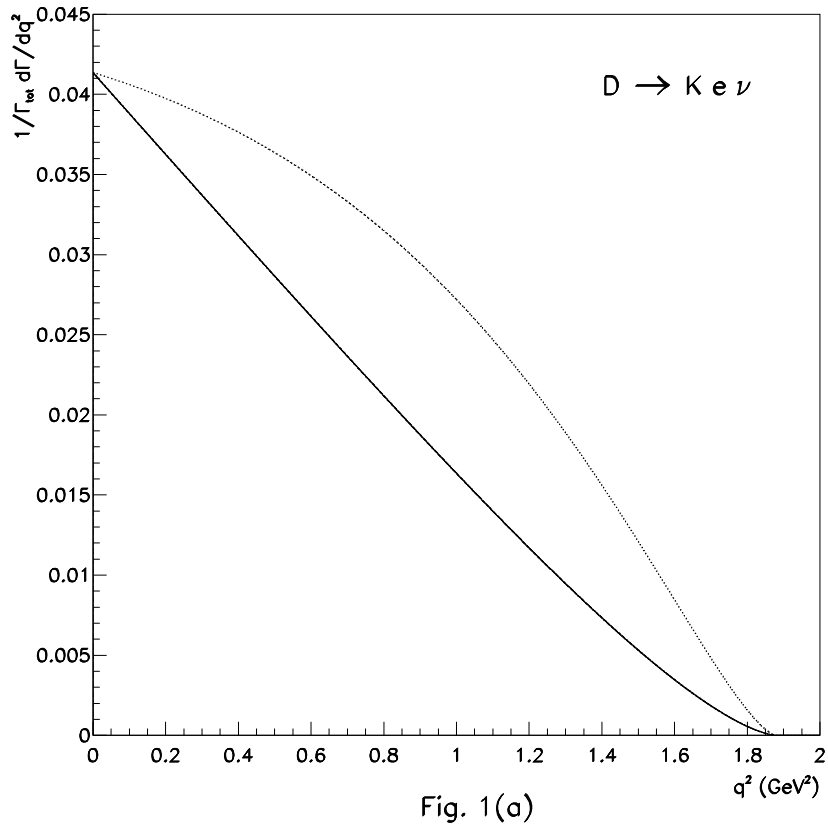
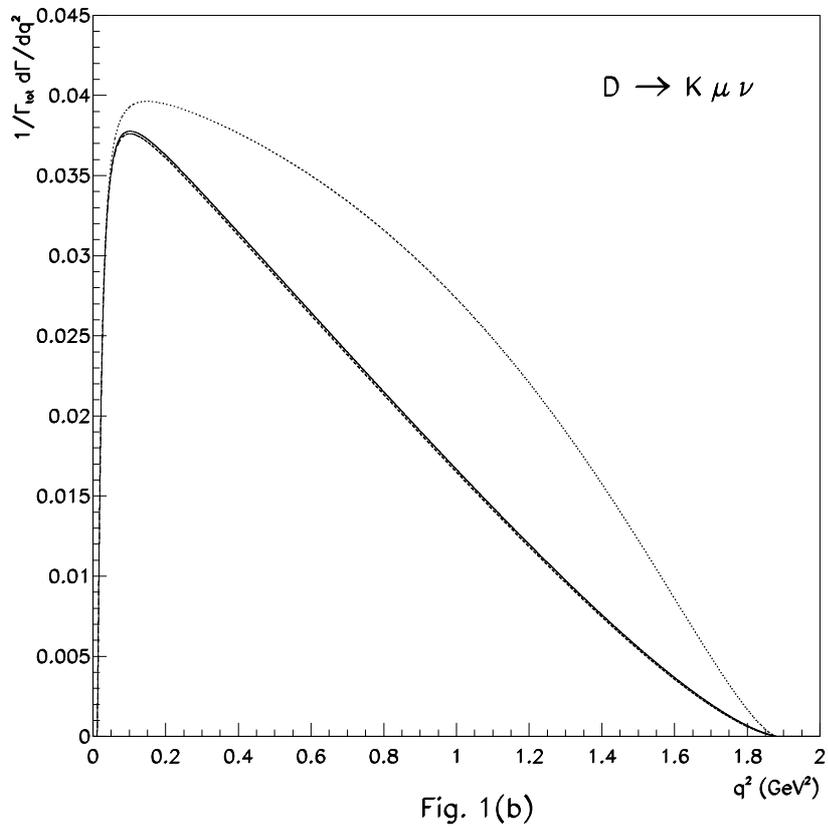
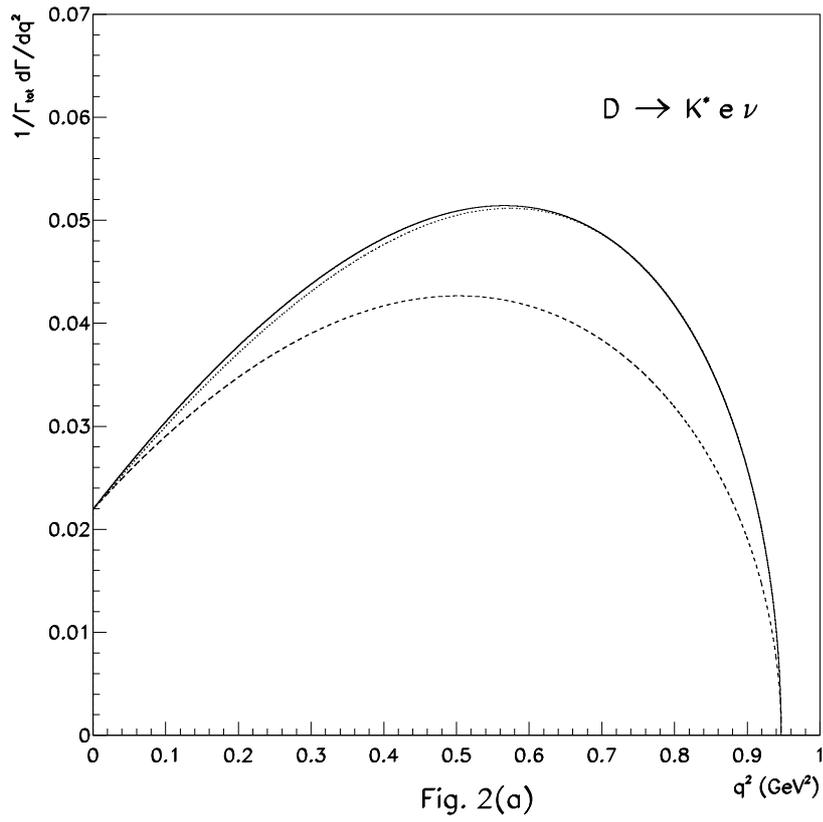
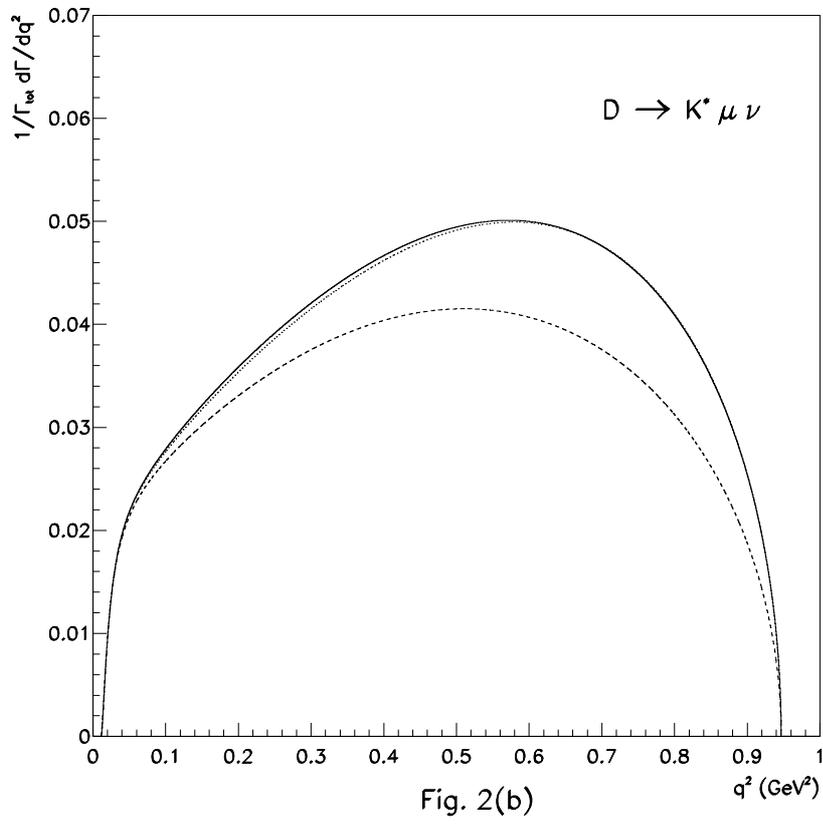
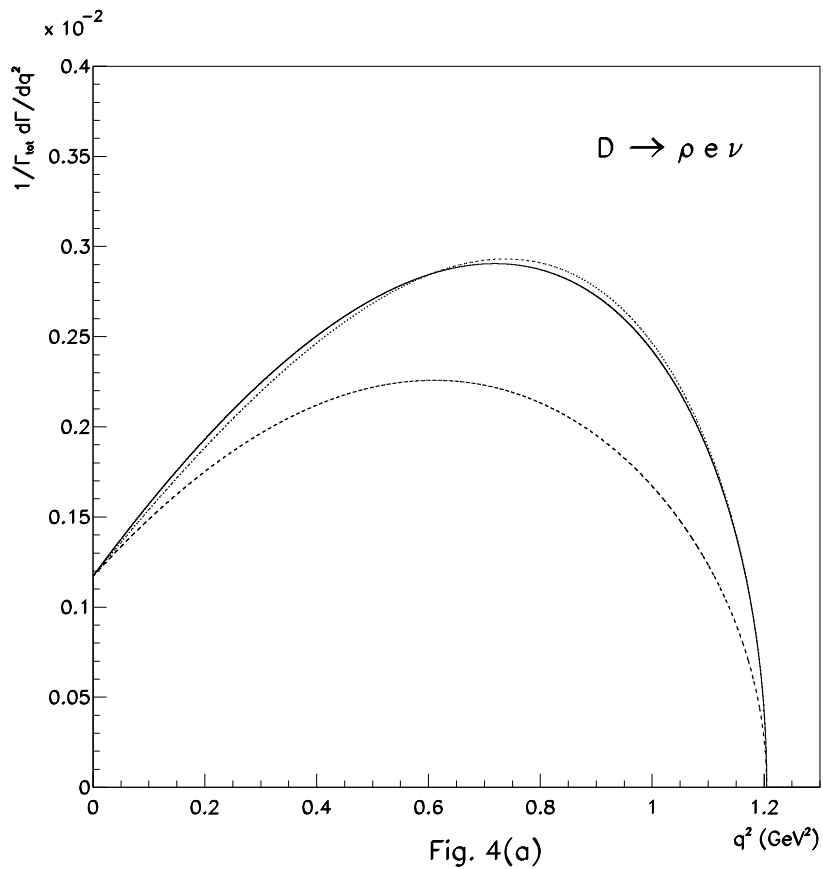
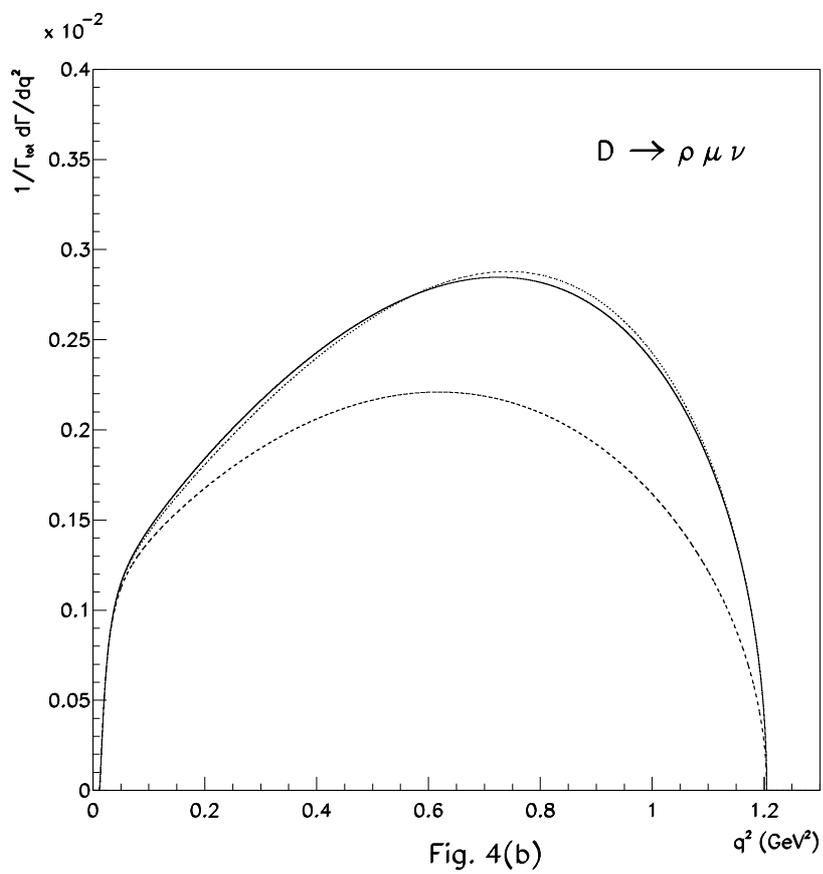
\end{document}